\documentclass{article}
\usepackage{ijcai24}

\usepackage{times}
\usepackage{soul}
\usepackage{url}
\usepackage[hidelinks]{hyperref}
\usepackage[utf8]{inputenc}
\usepackage[small]{caption}
\usepackage{graphicx}
\usepackage{amsmath}
\usepackage{amsthm}
\usepackage{booktabs}
\usepackage{algorithm}
\usepackage{algorithmic}
\usepackage[switch]{lineno}
\usepackage{multirow}
 \usepackage{amssymb}
 \usepackage{tablefootnote}
 \usepackage{diagbox}
 \usepackage{enumitem}

\title{A Survey on Cross-Domain Sequential Recommendation}

\author{
Shu Chen$^1$
\and
Zitao Xu$^1$\and
Weike Pan$^1$ \footnote{Corresponding author}\and
Qiang Yang$^{2,}$$^3$\and
Zhong Ming$^{1,}$$^{4}$
\affiliations
$^{1}$College of Computer Science and Software Engineering, Shenzhen University, P.R. China\\
$^{2}$WeBank AI Lab, WeBank, P.R. China\\
$^{3}$The Hong Kong University of Science and Technology, P.R. China\\
$^{4}$College of Big Data and Internet, Shenzhen Technology University, P.R. China
\emails
\{chenshu20, xuzitao2018\}@email.szu.edu.cn, \{panweike,mingz\}@szu.edu.cn, qyang@cse.ust.hk
}

\begin{document}

\maketitle

\begin{abstract}
Cross-domain sequential recommendation (CDSR) shifts the modeling of user preferences from flat to stereoscopic by integrating and learning interaction information from multiple domains at different granularities (ranging from inter-sequence to intra-sequence and from single-domain to cross-domain).
In this survey, we first define the CDSR problem using a four-dimensional tensor and then analyze its multi-type input representations under multidirectional dimensionality reductions. 
Following that, we provide a systematic overview from both macro and micro views.
 From a macro view, we abstract the multi-level fusion structures of various models across domains and discuss their bridges for fusion. 
From a micro view, focusing on the existing models, we first discuss the basic technologies and then explain the auxiliary learning technologies.
Finally, we exhibit the available public datasets and the representative experimental results as well as provide some insights into future directions for research in CDSR.
\end{abstract}

\section{Introduction}
In real life, people always leave interaction traces in multiple scenarios (e.g., shopping scenarios, reading scenarios, etc.), multiple platforms (e.g., Amazon, Taobao, etc.), and even multiple boards (e.g., books and movies in Douban, etc.).
Considering all these as different domains, combining in-depth information across domains is a critical step for the development of recommender systems.
It makes the data of user preferences no longer fragmented and shifts the modeling from flat to stereoscopic.
As shown in Table~\ref{tab:introduction}, recommender systems focus on information that is becoming deeper and more comprehensive, starting from traditional one-class collaborative filtering (OCCF), further progressing to sequential one-class collaborative filtering (SOCCF) and to cross-domain one-class collaborative filtering (CD-OCCF), and finally to cross-domain sequential one-class collaborative filtering (CD-SOCCF, also called cross-domain sequential recommendation, CDSR). 
\begin{table}[t]
\caption{A table summarizing OCCF, SOCCF, CD-OCCF and CD-SOCCF (CDSR) based on information from different granularity.}
\label{tab:introduction}
\resizebox{\linewidth}{!}{
\begin{tabular}{c|cc|cc}
\hline
\multirow{2}{*}{\diagbox[height=3.17em]{Problem}{Information}} & \multicolumn{2}{c|}{Single-Domain} & \multicolumn{2}{c}{Cross-Domain} \\ \cline{2-5} 
 & \multicolumn{1}{c|}{\begin{tabular}[c]{@{}c@{}}Inter-\\ sequence\end{tabular}} & \begin{tabular}[c]{@{}c@{}}Intra-\\ sequence\end{tabular} & \multicolumn{1}{c|}{\begin{tabular}[c]{@{}c@{}}Inter-\\ sequence\end{tabular}} & \begin{tabular}[c]{@{}c@{}}Intra-\\ sequence\end{tabular} \\ \hline
OCCF & \multicolumn{1}{c|}{{\checkmark}} &  & \multicolumn{1}{c|}{} &  \\ \hline
SOCCF & \multicolumn{1}{c|}{{\checkmark}} & {\checkmark} & \multicolumn{1}{c|}{} &  \\ \hline
CD-OCCF & \multicolumn{1}{c|}{{\checkmark}} &  & \multicolumn{1}{c|}{{\checkmark}} &  \\ \hline
CD-SOCCF (a.k.a. CDSR) & \multicolumn{1}{c|}{{\checkmark}} & {\checkmark} & \multicolumn{1}{c|}{{\checkmark}} & {\checkmark} \\ \hline
\end{tabular}
}
\end{table}
\begin{figure}
    \centering
\includegraphics[width=1.0\columnwidth]{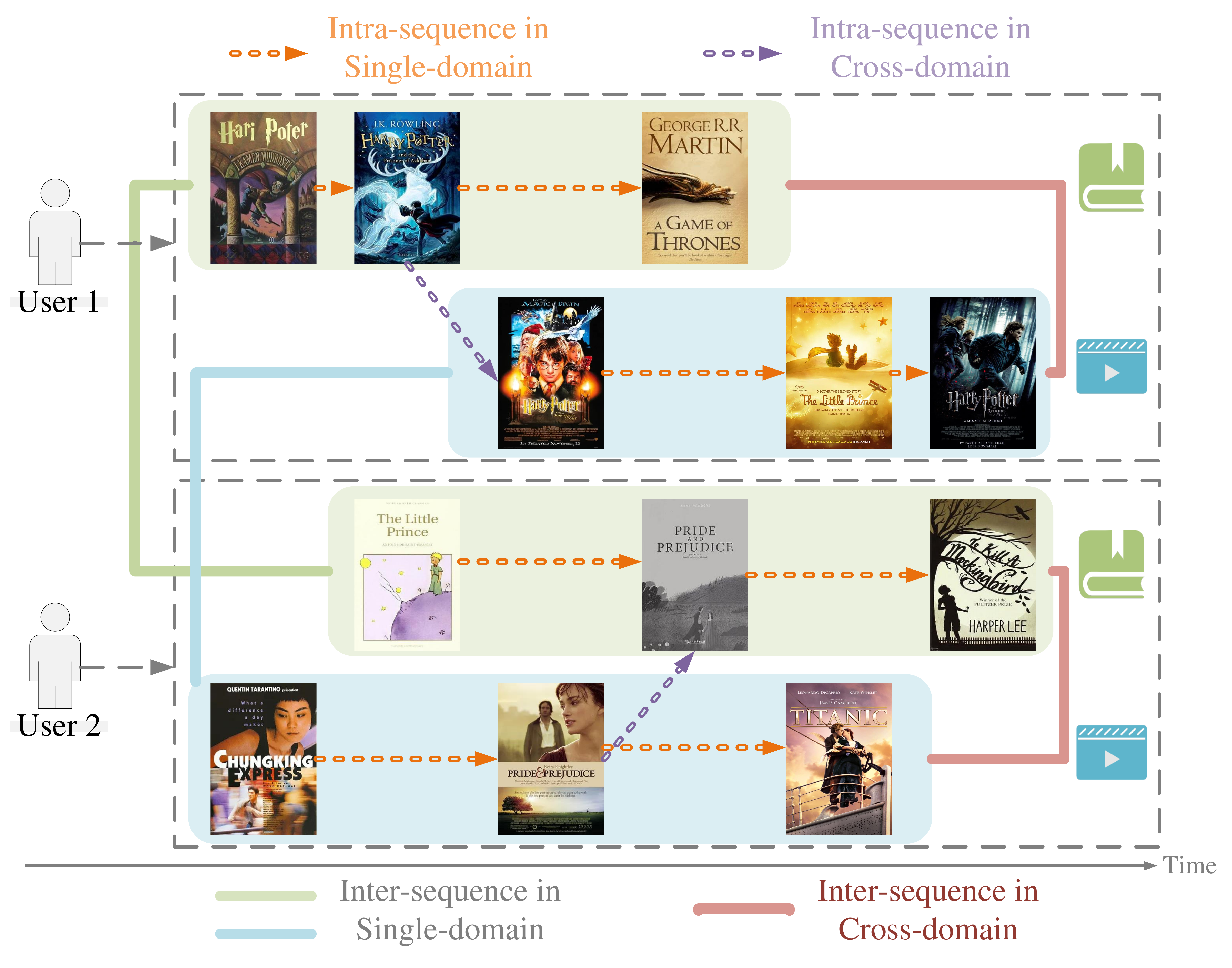}
\vspace{-1.5em}
    \caption{Illustration of CDSR.}
    \label{fig:CDSR}
\end{figure}

In the case of OCCF~\cite{BPR}, the available information is limited to whether a user interacts with an item. 
As illustrated in Figure~\ref{fig:CDSR}, we only know all the items that User 1 and User 2 have interacted with in a single domain, without considering the sequential pattern among successive items.
SOCCF~\cite{FISSA}, building upon OCCF, places greater emphasis on the order relationships between items within a sequence, i.e., the information of intra-sequence in a single domain, which enables the model to capture the user's long-term and short-term preferences.
At a large granularity, expanding from a single domain to multiple domains offers a fresh perspective to address the data sparsity issue in recommender systems, thus constituting CD-OCCF~\cite{CoNet}.
For instance, as shown in Figure~\ref{fig:CDSR}, users have interactions not only in the book domain but also in the closely related movie domain. 
In a cross-domain scenario, we must meet the following challenges: how to transfer or aggregate interaction records from multiple domains and how to align representations of users or items across different domains~\cite{Cross-domain-survey}.

Indeed, CDSR combines all of the aforementioned directions and further incorporates sequential information based on CD-OCCF and cross-domain information based on SOCCF.
As indicated by the purple arrows in Figure~\ref{fig:CDSR}, CDSR necessitates the precise capturing of sequential relationships between items across different domains. 
For example, User 1, after reading two books in the $Harry\ Potter$ series, chooses to watch the corresponding $Harry\ Potter$ movie. 
Similarly, User 2 who enjoys watching romantic movies also reads the original novels after watching the $Pride\ and\ Prejudice$ movie.
Besides, it is also worth considering in CDSR how to fuse sequential information from different domains as well as how to distinguish between users' specific preferences within a single domain and global preferences shared across multiple domains.

In this survey, we first formulate the CDSR problem and modeling tasks,  considering various dimensionality reductions and different input representations.
Then we adopt both macro and micro views to summarize the existing works in CDSR. 
From a macro view, we present an overview of multi-level fusion structures, discussing how to fuse information across different domains and exploring bridges for cross-domain fusion. 
From a micro view, we conduct a detailed analysis of various technologies employed by existing works that are categorized into basic and auxiliary learning technologies.
Based on this analysis, we obtain a comprehensive classification (shown in Table~\ref{tab:total}) and framework figure of the key technologies (shown in Figure~\ref{fig:framework}).
Furthermore, we list the datasets commonly used in CDSR and the representative experimental results as well as provide some insights into potential future directions.

\section{Formulation}
In this section, we first present the problem definition of CDSR using a four-dimensional tensor. 
Then, referring to Figure~\ref{fig:tensor}, we delve into the various directions of dimensionality reduction that are often adopted in practical modeling and the corresponding multi-type input representations.
\subsection{Problem Definition}
\label{sec:ProblemDefinition}
As temporal information and cross-domain information are incorporated, the interaction data between users and items in CDSR undergoes a gradual expansion into a four-dimensional data tensor, consisting of dimensions about users, items, time, and domains.
We denote this four-dimensional tensor with $\Gamma \in \mathbb{R}^{n\times m\times s\times k}$, where $n$ and $m$ is the number of users and items in all domain, $s$ is the discrete time intervals, and $k$ is the number of domains.
Each element $\gamma_{(u,i,t,d)}$ in it signifies whether there is an interaction between user $u$ and item $i$ at time $t$ in domain $d$.
The goal of CDSR is to estimate the probability for all candidate items in each domain, and then to recommend the most probable next item to each user. 
The estimated probability can be formalized as follows,
\begin{align}
\label{eq:problemDefiniton}
            P(\ \hat{i}\ |\ \Gamma\ )\sim f(\ \Gamma\ )
\end{align}
where $\hat{i}$ denotes the candidate item of each user in each domain, $\Gamma$ is the four-dimensional tensor containing all interaction information, and $f(\ \Gamma\ )$ indicates the learned function to estimate $P(\ \hat{i}\ |\ \Gamma\ )$.

Notice that most existing works in CDSR are primarily established on two domains, i.e., $k=2$. Therefore, for the subsequent discussion, we take two domains as an example and represent them as domain A and domain B, respectively.
In fact, some researchers~\cite{DCDIR,MGCL} also refer to the two domains as a source domain and a target domain.  
Moreover, some works~\cite{pi-Net,PSJNet,DA-GCN,RL-ISN} introduce an additional assumption with a shared account. They assume that within a real account, there are $q$ virtual users ($u_j, j \leq q$) simultaneously active.
Regarding the shared account issue, it is applicable not only in cross-domain scenarios but also in single-domain scenarios.

\subsection{Multidirectional Dimensionality Reduction}
\begin{figure}
    \centering
    \includegraphics[width=1.0\columnwidth]{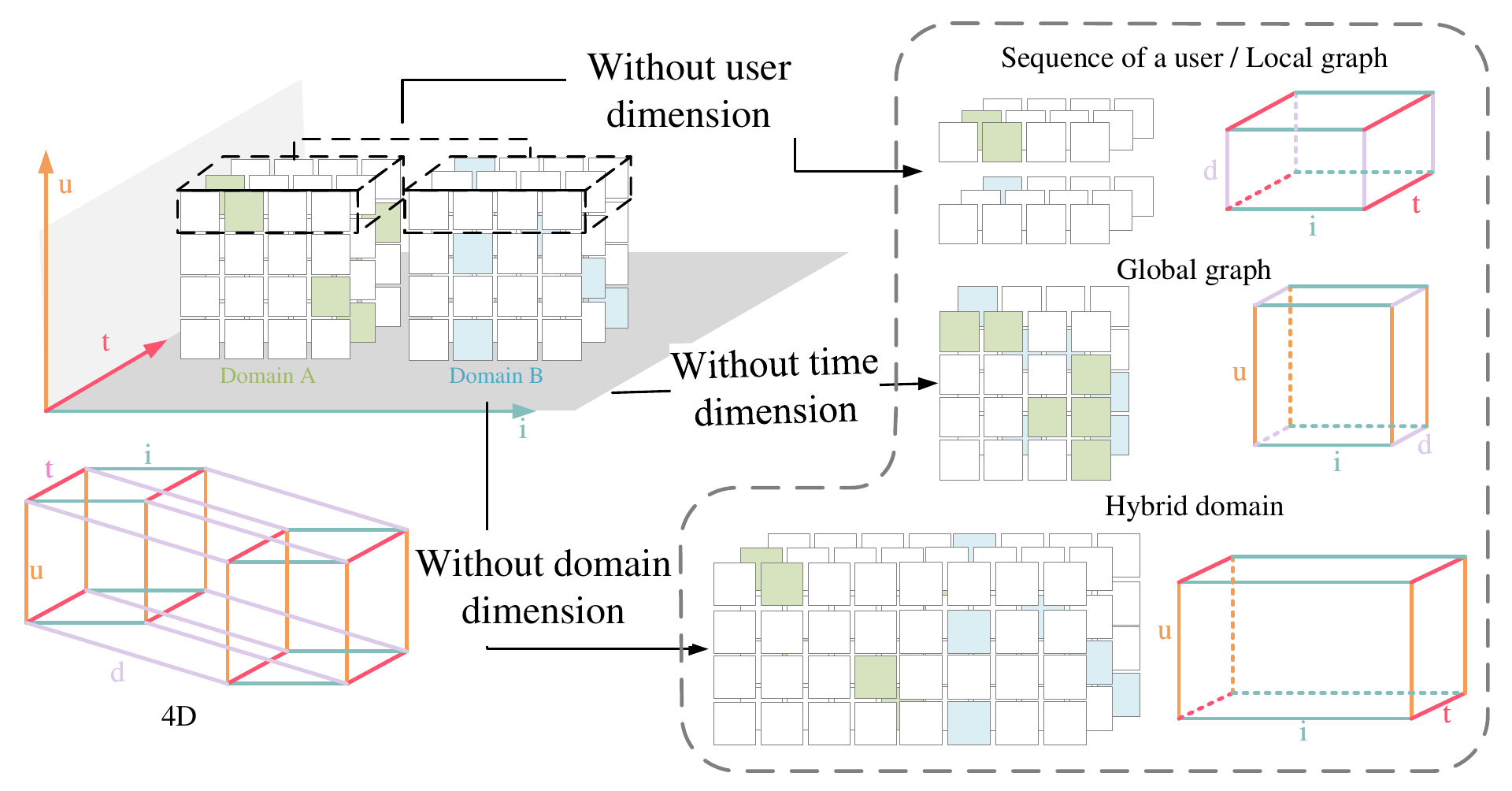}
    \vspace{-2em}
    \caption{A visualization of dimensionality reduction for a four-dimensional data tensor in CDSR scenarios. We represent the dimensions contained in the data using wireframes and simulate tensors using blocks. The colored blocks indicate records of user interaction with items within the domain at a given moment. In this case, both domains share the same set of users, but there is no overlap in the items they interact with.}
    \label{fig:tensor}
\end{figure}
For computational and modeling convenience, the four-dimensional tensor is often reduced to some three-dimensional tensors from various directions (as shown in Figure~\ref{fig:tensor}). Below, we explain dimensionality reduction from three directions:
\begin{itemize}[leftmargin=*]
    \item \textbf{W/o User Dimension.} For a user, his or her interactions in chronological order can form a sequence in each domain.  These sequences reflect the user's personalization as well as long and short-term preferences in different domains, and can further serve as the basis for constructing a local graph in subsequent work. 
    \item \textbf{W/o Time Dimension.} Mapping the tensor along the temporal dimension and aggregating interactions within a time lays the groundwork for a global graph. It allows models to learn user-to-user and user-to-item connections, which are more skewed toward users' global preferences.
    \item \textbf{W/o Domain Dimension.} Blending two or more domains into a hybrid domain, a model can clarify the passing relationships between items originally in different domains.  
    Many existing studies treat the hybrid domain as an independent domain, and design model fusion structures that incorporate all three domains in parallel, which is shown in Section~\ref{sec:inter-domain}.
\end{itemize}

 Indeed, these dimensionality reduction directions are often combined and used in conjunction rather than being treated as independent.
 Adopting only one reduction direction may result in missing information to a certain degree.
 Next, we visualize these reduction processes on different input representations.

\subsection{Multi-Type Input Representations}
\label{sec:Intra-domain}
Researchers' considerations on dimensionality reduction directions in CDSR are first reflected in the input representations. 
Here, we categorize conventional input representations into pure sequential representation and graph-encoded sequential representation. 
Subsequently, we analyze unconventional inputs, including side information and pre-trained features.

\subsubsection{Pure Sequential Representation}

Arranging the interacted items of a user in chronological order yields a sequence.
For each user $u$, we define $S^A_u=\{i^A_1, i^A_2, \dots\}$ and $S^B_u=\{i^B_1, i^B_2, \dots\}$ as his or her interaction sequences in domain A and domain B, respectively.
If $u$ is an overlapping user who has interactions both in domain A and domain B, we can mix $S^A_u$ and $S^B_u$ to a hybrid sequence in chronological order (e.g., $S^{hybrid}_u=\{i^A_1, i^B_1, i^A_2, i^A_3, i^B_2,\dots\}$), which is viewed as a sequence of the hybrid domain.
From a pure sequential perspective, we can substitute the four-dimensional tensor in Eq.(\ref{eq:problemDefiniton}) with the set of sequences in each domain, as follows,
\begin{align}
     P(\ \hat{i}\ |S^A,S^B)\sim f(S^A,S^B)
\end{align}

\subsubsection{Graph-Encoded Sequential Representation}
Some researchers~\cite{DA-GCN,DDGHM,C2DSR,LEA-GCN} turn to construct directed graphs $G=\{V,E\}$ to model sequential information, where $V$ is a set of items that have been interacted with and $E$ is the edges that represent relations of the serial relationship from item to item. 
Constructing a user's sequence within a domain or a sequence within a session as a local graph is a common practice~\cite{DDGHM,DAT-MDI}.
In the global graph construction scenario, $E$ is also utilized to denote the relations between users and items~\cite{MGCL}.
We represent the raw data in its graph-encoded form to extend Eq.(\ref{eq:problemDefiniton}), where $G_l$ denotes all local graphs and $G_g$ denotes the global graph, as follows,
\begin{align}
     P(\ \hat{i}\ |G_l,G_g)\sim f(G_l,G_g)
\end{align}

\subsubsection{Side Information}
More and more researchers consider incorporating side information to enrich the semantic representation of users' historical behaviors. 
In this section, we divide the most common side information into three categories: time, text (e.g., user/item contexts and reviews), and knowledge graph.

\textbf{Time.}
Previously the timestamps are only used to order the items, but some researchers~\cite{P-CDSR,CsrGCF,TiDA-GCN} define $t_{ij}=|t_i-t_j|$ to model the time interval and apply it to the subsequent learning of model, where $t_i$ and $t_j$ are the timestamps that the item $i$ and the item $j$ are interacted with, respectively.
Besides, we can also mine more information about time, such as periodicity and the duration of the interaction, etc.

\textbf{User/Item Contexts and Reviews.}
Except for the user ID, triples composed of multiple contextual information (e.g., $(``User ID",``City", ``Age", \cdots)$) are utilized as a basis for finding user-user relationships~\cite{MiNet}.
Similarly, the context information that consists of categories, tags, keywords, etc, is used as a supplement to item ID~\cite{P-CDSR,MiNet,CDNST}.
Review as a type of available text message can associate users and items well.
A common way to encode these texts is through some pre-trained models, such as BERT~\cite{BERT}.

\textbf{Knowledge Graph.}
Knowledge graphs~\cite{DCDIR,MIFN} deal with information from both entity and relationship views.
It not only contains structural information between nodes but also implies some relationship description.
A knowledge graph defines an entity set $E^{KG}$ and a relation set $R^{KG}$, which consists of multiple entity-relation-entity triples $<e_i, r, e_j>$ (e.g., $<e_{i_1^A}, Is\_the\_same\_category,e_{i_1^B}>$ meaning that the entity $e_{i_1^A}$ from domain A has the same category as entity $e_{i_1^B}$ from domain B).

Considering the aforementioned side information, we can extend Eq.(\ref{eq:problemDefiniton}) as follows,
\begin{align}
     P(\ \hat{i}\ |\ \Gamma\ ,D)\sim f(\ \Gamma\ ,D )
\end{align}
where $D$ is the side information mentioned above and often serves as supplementary data to conventional inputs.

\subsubsection{Pre-trained Features}
Considering the privacy issues in two or more domains, some works~\cite{TPUF,FedDCSR,SEMI} use model-trained features as the input from the source domain, instead of raw data.
\cite{TPUF} directly combines the pre-trained user features from one domain with another domain.
\cite{FedDCSR} utilizes federated learning to fetch the global representation from the server, and then injects it into the local model.

We take domain A as an example and revise Eq.(\ref{eq:problemDefiniton}) as follows,
\begin{align}
    P(\ \hat{i}\ |\ \Gamma^A\ ,M^B)\sim f(\ \Gamma^A\ ,M^B)
\end{align}
where $M^B$ are the features of domain B after pre-training and $\Gamma^A$ denotes the interaction data exclusively within domain A. 
Notice that in domain B, the equation is $P(\ \hat{i}\ |\ \Gamma^B\ ,M^A)\sim f(\ \Gamma^B\ ,M^A)$.

\section{Macro-View: What Structure Is Used to Fuse the CDSR Information?}
In CDSR, the fusion structure serves as the primary skeleton of a model, providing pathways for the separation and aggregation of these features.
In this section, we first describe multi-level fusion structures and then elaborate on the bridge for inter-domain fusion.

\begin{figure}
    \centering
    \includegraphics[width=0.92\columnwidth]{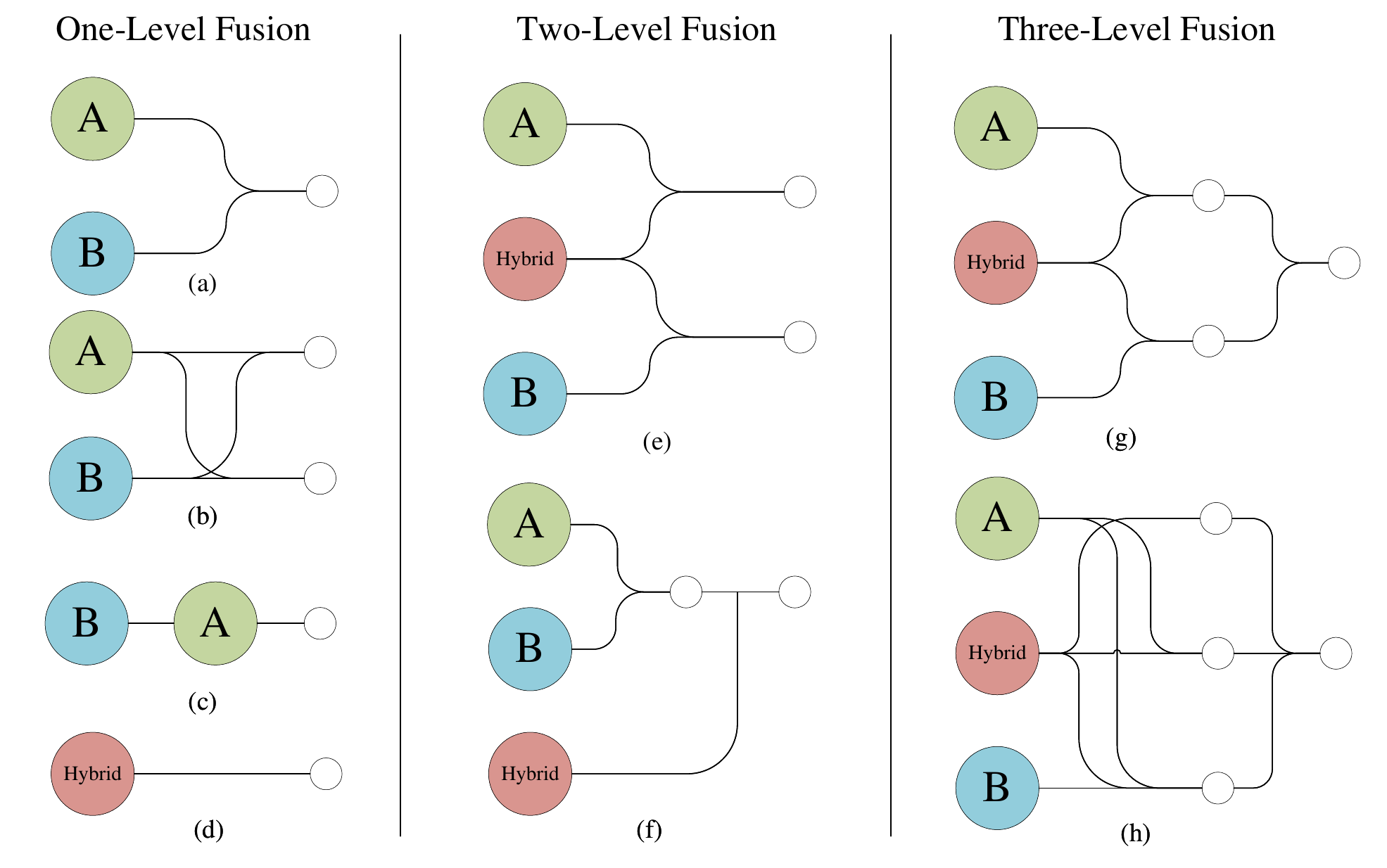}
    \vspace{-0.5em}
    \caption{The overview of multi-level fusion structures that are divided into three levels. ``A'' and ``B'' represent domain A and domain B, respectively, and ``Hybrid'' denotes a combination of two domains in chronological order.}
    \label{fig:structure}
\end{figure}

\subsection{Multi-Level Cross-Domain Fusion Structures}
\label{sec:inter-domain}
For cross-domain fusion, researchers employ various structures to aggregate information from different domains.
In this section, we overview the various multi-level fusion structures shown in Figure~\ref{fig:structure}. 
Specifically, we introduce those structures from two parts, i.e., the one-level fusion structures and the two/three-level fusion structures.

\subsubsection{One-Level Fusion Structures}
In earlier works, to combine the cross-domain information, there are a lot of works that first learn user preferences from domain A and domain B separately, and then fuse the representations of the two domains using various operations (as shown in Figure~\ref{fig:structure}(a)). 
These operations include, but are not limited to,  concatenation~\cite{SEMI,RL-ISN}, summation~\cite{CD-ASR,TPUF}, multi-layer perceptron (MLP)~\cite{DCDIR}, and some attention mechanisms~\cite{MiNet,DASL}.

Additionally, some works employ transfer learning to transfer knowledge from a source domain to a target domain~\cite{DAT-MDI,SATLR}, or train a discriminator to bridge the representations of two domains with the idea of adversarial learning~\cite{RecGURU}, as shown in Figure~\ref{fig:structure}(b).

In contrast to the above-juxtaposed structures, there are also works utilizing a tandem structure to fuse information~\cite{CD-SASRec} (i.e., Figure~\ref{fig:structure}(c)), or directly tackling the problem from the perspective of a hybrid-domain view (i.e., Figure~\ref{fig:structure}(d)).
For instance, some works~\cite{DA-GCN,TiDA-GCN} construct a global graph for the hybrid domain to combine the cross-domain knowledge.
And others like~\cite{pi-Net,PSJNet} choose to learn the item transition patterns in the hybrid sequences $S^{hybrid}$.

\subsubsection{Two/Three-Level Fusion Structures}
Considering the potential for further exploration and advancement in fusion structures, some works combine the single domain and the hybrid domain with a multi-level fusion structure.
As shown in Figure~\ref{fig:structure}(e), in order to predict the next item in a separate domain, the hybrid domain is utilized as the main sharer~\cite{C2DSR,CsrGCF} or the bridge~\cite{DDGHM} to transfer knowledge from another domain. 
There are also some works~\cite{DREAM} that choose to combine the domain A and domain B first and then fuse the hybrid information (i.e., Figure~\ref{fig:structure}(f)).

To delve further into the fusion structures, some researchers continue to extend the hierarchy, as illustrated in Figure~\ref{fig:structure}(g)~\cite{MGCL} which aggregates the representations again after combining the hybrid domain information on the basis of Figure~\ref{fig:structure}(e).
Figure~\ref{fig:structure}(h)~\cite{LEA-GCN} proposes a more complex structure, which shares the coarse-grained representations of the target domain A and the hybrid domain with each other domain.
\subsubsection{Discussion}
Although there are various multi-level fusion structures, it does not mean that a more complex structure will perform better.
Limited by the degree of fusion, the simple fusion structures, i.e., one-level fusion structures, are easy to implement but may not comprehensively model the domain-specific and domain-generic features.
While improving effectiveness, multi-level fusion structures bring increased complexity and reduced interpretability.

Considering the symmetry and asymmetry of the structures, which the researchers focus on each domain greatly affects the design of a cross-domain fusion structure. 
For example, some researchers~\cite{MiNet,DCDIR,MGCL} consider the two domains as source and target domains and leverage the data-rich source domain to assist the data-sparse target domain, which makes the target domain dominant in the fusion structure. 
Other researchers~\cite{DA-GCN,DASL,MIFN} aim to improve the recommendation performance of both domains simultaneously, making the fusion structure designed tend to be more symmetric.
\begin{figure}
    \centering
    \includegraphics[width=1.0\columnwidth]{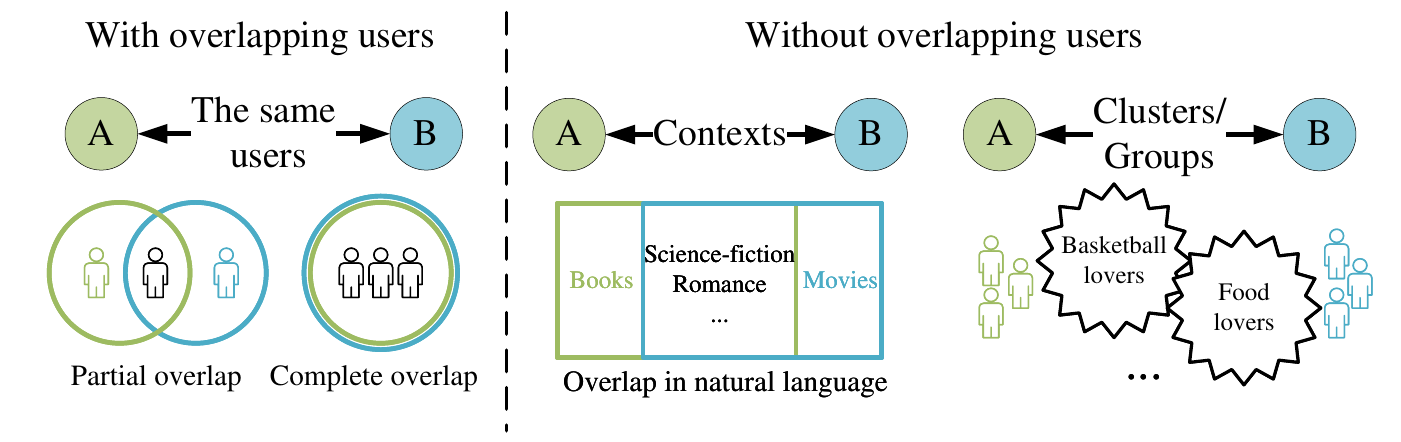}
    \vspace{-2em}
    \caption{Examples of building cross-domain bridges relying on different information.}
    \label{fig:overlap}
\end{figure}
\begin{table*}[t]
\caption{A systematic overview of the existing models for CDSR.}
\label{tab:total}
\centering
\resizebox{0.95\textwidth}{!}{
\begin{tabular}{c|ccc|ccc|l}
\hline
\multirow{3}{*}{Data   Structure} & \multicolumn{3}{c|}{Basic   Technology} & \multicolumn{3}{c|}{Auxiliary   Learning} & \multicolumn{1}{c}{\multirow{3}{*}{Paper}} \\ \cline{2-7}
 & \multicolumn{1}{c|}{GNN} & \multicolumn{1}{c|}{RNN} & Attention & \multicolumn{1}{c|}{\begin{tabular}[c]{@{}c@{}}Contrastive\\ Learning\end{tabular}} & \multicolumn{1}{c|}{\begin{tabular}[c]{@{}c@{}}Transfer\\ Learning\end{tabular}} & Others & \multicolumn{1}{c}{} \\ \hline
\multirow{18}{*}{Sequence} & \multicolumn{1}{c|}{\multirow{11}{*}{}} & \multicolumn{1}{c|}{\multirow{4}{*}{{\checkmark}}} & \multirow{2}{*}{} & \multicolumn{1}{c|}{\multirow{3}{*}{}} & \multicolumn{1}{c|}{} & \multirow{7}{*}{} & \begin{tabular}[c]{@{}l@{}}$\pi$-Net~\cite{pi-Net},  CDHRM~\cite{CDHRM},\\SCLSTM~\cite{SCLSTM}, PSJNet~\cite{PSJNet}\end{tabular} \\ \cline{6-6} \cline{8-8} 
 & \multicolumn{1}{c|}{} & \multicolumn{1}{c|}{} &  & \multicolumn{1}{c|}{} & \multicolumn{1}{c|}{\multirow{2}{*}{{\checkmark}}} &  & CDNST~\cite{CDNST} \\ \cline{4-4} \cline{8-8} 
 & \multicolumn{1}{c|}{} & \multicolumn{1}{c|}{} & \multirow{14}{*}{{\checkmark}} & \multicolumn{1}{c|}{} & \multicolumn{1}{c|}{} &  & DASL~\cite{DASL}, TJAPL~\cite{TJAPL} \\ \cline{5-6} \cline{8-8} 
 & \multicolumn{1}{c|}{} & \multicolumn{1}{c|}{} &  & \multicolumn{1}{c|}{\multirow{4}{*}{{\checkmark}}} & \multicolumn{1}{c|}{\multirow{3}{*}{}} &  & CMVCDR~\cite{CMVCDR} \\ \cline{3-3} \cline{8-8} 
 & \multicolumn{1}{c|}{} & \multicolumn{1}{c|}{\multirow{7}{*}{}} &  & \multicolumn{1}{c|}{} & \multicolumn{1}{c|}{} &  & \begin{tabular}[c]{@{}l@{}}SEMI~\cite{SEMI}, DREAM~\cite{DREAM}, P-CDSR~\cite{P-CDSR},\\ Tri-CDR~\cite{Tri-CDR}, CGRec~\cite{CGRec}, LCN~\cite{LCN},\\ MACD~\cite{MACD}\end{tabular} \\ \cline{5-5} \cline{8-8} 
 & \multicolumn{1}{c|}{} & \multicolumn{1}{c|}{} &  & \multicolumn{1}{c|}{\multirow{6}{*}{}} & \multicolumn{1}{c|}{} &  & \begin{tabular}[c]{@{}l@{}}MiNet~\cite{MiNet}, CD-ASR~\cite{CD-ASR},\\ CD-SASRec~\cite{CD-SASRec}, MAN~\cite{MAN}\end{tabular} \\ \cline{6-6} \cline{8-8} 
 & \multicolumn{1}{c|}{} & \multicolumn{1}{c|}{} &  & \multicolumn{1}{c|}{} & \multicolumn{1}{c|}{{\checkmark}} &  & SATLR~\cite{SATLR} \\ \cline{6-8} 
 & \multicolumn{1}{c|}{} & \multicolumn{1}{c|}{} &  & \multicolumn{1}{c|}{} & \multicolumn{1}{c|}{\multirow{4}{*}{}} & \begin{tabular}[c]{@{}c@{}}Adversarial\\ Learning\end{tabular} & RecGURU~\cite{RecGURU}, TPUF~\cite{TPUF}, DA-DAN~\cite{DA-DAN} \\ \cline{7-8} 
 & \multicolumn{1}{c|}{} & \multicolumn{1}{c|}{} &  & \multicolumn{1}{c|}{} & \multicolumn{1}{c|}{} & \begin{tabular}[c]{@{}c@{}}Reinforcement\\ Learning\end{tabular} & RL-ISN~\cite{RL-ISN}, O-SCDR~\cite{O-SCDR} \\ \cline{7-8} 
 & \multicolumn{1}{c|}{} & \multicolumn{1}{c|}{} &  & \multicolumn{1}{c|}{} & \multicolumn{1}{c|}{} & \begin{tabular}[c]{@{}c@{}}Prompt\\ Learning\end{tabular} & PLCR~\cite{PLCR} \\ \cline{4-4} \cline{7-8} 
 & \multicolumn{1}{c|}{} & \multicolumn{1}{c|}{} &  & \multicolumn{1}{c|}{} & \multicolumn{1}{c|}{} &  & MSECDR~\cite{MSECDR}, AMID~\cite{AMID} \\ \hline
\multirow{10}{*}{Graph} & \multicolumn{1}{c|}{\multirow{10}{*}{{\checkmark}}} & \multicolumn{1}{c|}{\multirow{3}{*}{{\checkmark}}} &  & \multicolumn{1}{c|}{\multirow{4}{*}{}} & \multicolumn{1}{c|}{\multirow{2}{*}{}} & \multirow{5}{*}{} & DCDIR~\cite{DCDIR}, MIFN~\cite{MIFN} \\ \cline{4-4} \cline{8-8} 
 & \multicolumn{1}{c|}{} & \multicolumn{1}{c|}{} & \multirow{8}{*}{{\checkmark}} & \multicolumn{1}{c|}{} & \multicolumn{1}{c|}{} &  & AGNNGRU-CDR~\cite{AGNNGRU-CDR} \\ \cline{6-6} \cline{8-8} 
 & \multicolumn{1}{c|}{} & \multicolumn{1}{c|}{} &  & \multicolumn{1}{c|}{} & \multicolumn{1}{c|}{{\checkmark}} &  & DAT-MDI~\cite{DAT-MDI}, SGCross~\cite{SGCross} \\ \cline{3-3} \cline{6-6} \cline{8-8} 
 & \multicolumn{1}{c|}{} & \multicolumn{1}{c|}{\multirow{6}{*}{}} &  & \multicolumn{1}{c|}{} & \multicolumn{1}{c|}{\multirow{6}{*}{}} &  & DA-GCN~\cite{DA-GCN}, TiDA-GCN~\cite{TiDA-GCN} \\ \cline{5-5} \cline{8-8} 
 & \multicolumn{1}{c|}{} & \multicolumn{1}{c|}{} &  & \multicolumn{1}{c|}{\multirow{4}{*}{{\checkmark}}} & \multicolumn{1}{c|}{} &  & C$^2$DSR~\cite{C2DSR}, EA-GCL~\cite{EA-GCL}, MGCL~\cite{MGCL} \\ \cline{7-8} 
 & \multicolumn{1}{c|}{} & \multicolumn{1}{c|}{} &  & \multicolumn{1}{c|}{} & \multicolumn{1}{c|}{} & \begin{tabular}[c]{@{}c@{}}Federated\\ Learning\end{tabular} & FedDCSR~\cite{FedDCSR} \\ \cline{7-8} 
 & \multicolumn{1}{c|}{} & \multicolumn{1}{c|}{} &  & \multicolumn{1}{c|}{} & \multicolumn{1}{c|}{} & \multirow{3}{*}{} & DDGHM~\cite{DDGHM} \\ \cline{5-5} \cline{8-8} 
 & \multicolumn{1}{c|}{} & \multicolumn{1}{c|}{} &  & \multicolumn{1}{c|}{\multirow{2}{*}{}} & \multicolumn{1}{c|}{} &  & LEA-GCN~\cite{LEA-GCN}, IESRec~\cite{IESRec} \\ \cline{4-4} \cline{8-8} 
 & \multicolumn{1}{c|}{} & \multicolumn{1}{c|}{} &  & \multicolumn{1}{c|}{} & \multicolumn{1}{c|}{} &  & CsrGCF~\cite{CsrGCF} \\ \hline
\end{tabular}
}
\end{table*}
\subsection{Bridges for Cross-Domain Fusion}
When comes to cross-domain fusion, it is vital to clarify the bridges for information sharing between domains. 
We divide those bridges into three categories: same users, contexts, and clusters/groups, as shown in Figure~\ref{fig:overlap}.
The first is prevalent in scenarios with overlapping users, while the latter two are applied in scenarios without overlapping users.
\begin{itemize}[leftmargin=*]
    \item \textbf{Same Users.} 
    If some same users exist, then they are often the first choice as the bridge. Some works only focus on the fully overlapping users, neglecting non-overlapping users, while others~\cite{RecGURU} consider both cases.
    \item \textbf{Contexts.} If there are no common users, it is feasible to leverage the semantic similarity of natural language. For instance, we can replace traditional item IDs with item contexts to explore similarities between items of different domains~\cite{IESRec}.
    \item \textbf{Clusters/Groups.}  Further restricting the condition to no overlapping users and no side information, the clusters/groups are 
 another entry point since a specific group of users could have similar preferences~\cite{MAN}.
\end{itemize}

Works relying on overlapping users often perform well but their application scenarios may face limitations due to the sparse real-world data. 
Conversely, works built on non-overlapping users can be applied to a broader range of scenarios but may have limited performance.
In fact, cross-domain can be built on more than one type of bridge.

\section{Micro-View: What Technologies Are Used to Address the CDSR Problem?}
\label{sec:keyTechnologies}
In this section, we take a more tangible perspective to summarize the technologies adopted by the existing models in addressing the challenges in CDSR.
In conjunction with Table~\ref{tab:total} and Figure~\ref{fig:framework}, we elaborate on the basic technologies and the auxiliary learning technologies, respectively.
In Figure~\ref{fig:framework_example}, we give some examples of the utilization of the base technologies.
\subsection{Basic Technologies}
\begin{figure}[t]
    \centering
    \includegraphics[width=0.90\linewidth]{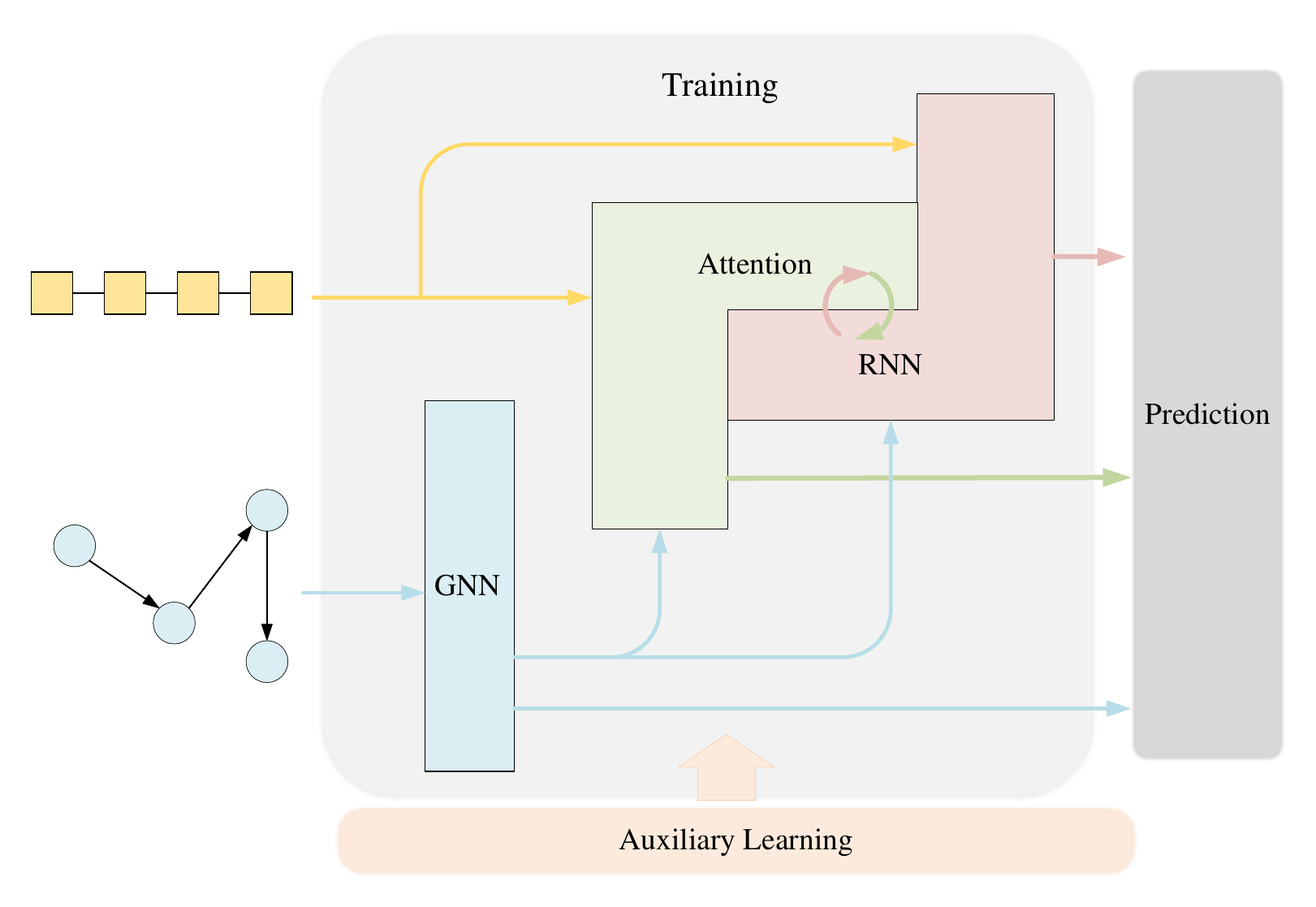}
    \caption{A schematic overview of the key technical framework. The color of the arrows represents the output after passing through the components represented by different colors. While models represented via graph structures require encoding with GNN, the relationship between RNN and attention can be used in parallel or an alternating fashion.}
    \label{fig:framework}
\end{figure}

\begin{figure}[t]
    \centering
    \includegraphics[width=1\linewidth]{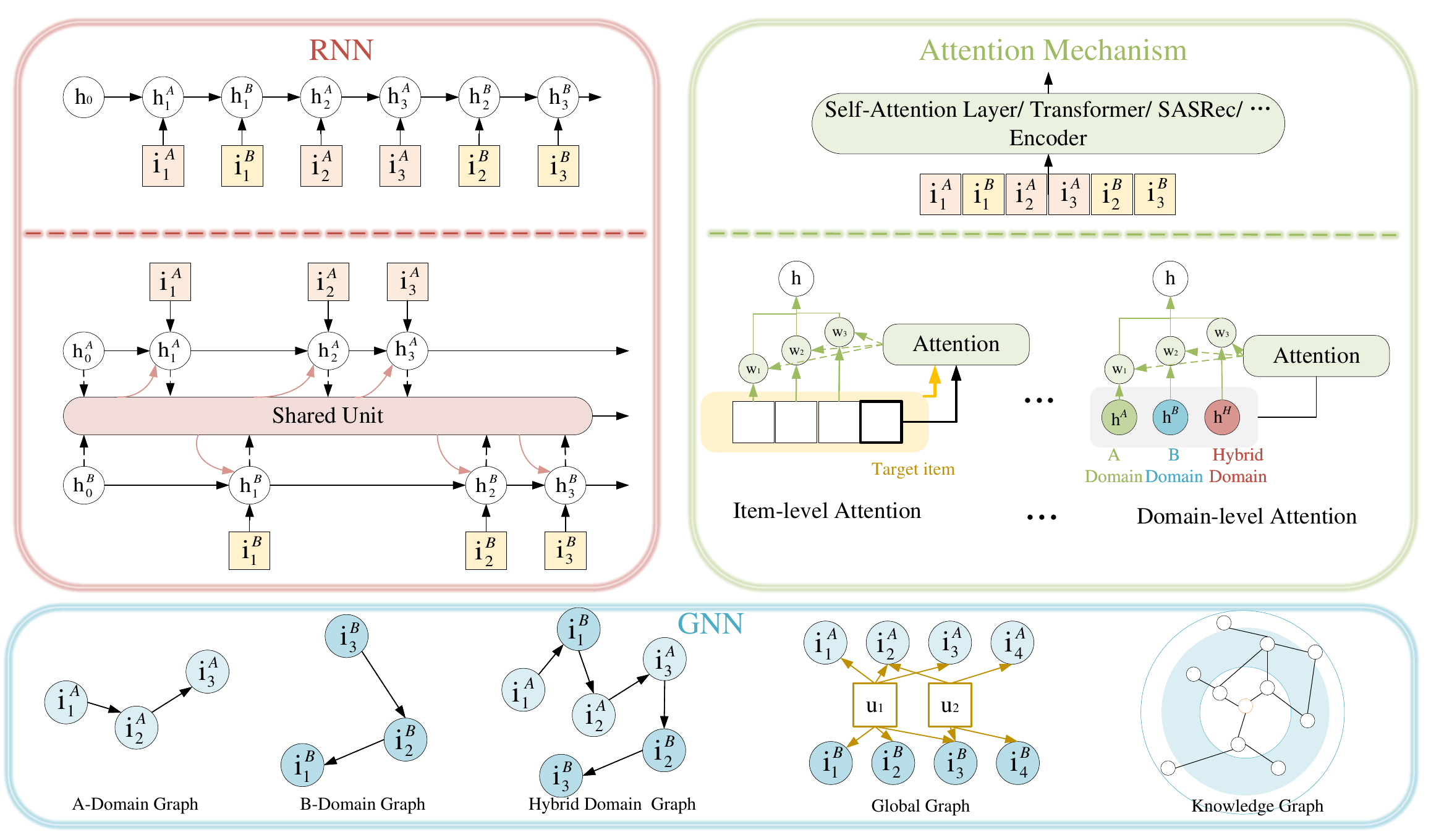}
    \caption{Some examples of basic technologies being applied to CDSR. $\{i^A_1,i^A_2,\cdots\}$ denotes the sequence of user interactions in domain A and similarly $\{i^B_1,i^B_2,\cdots\}$ denotes the sequence of interactions in domain B. $h$ denotes the output representation of the corresponding item after being processed by the model, and $w$ denotes the attention weight.}
    \label{fig:framework_example}
\end{figure}

According to Section~\ref{sec:Intra-domain}, the cross-domain information is always considered as a pure sequential representation or a graph-encoded sequential representation.
So we illustrate the utilization of sequence modeling technologies (i.e., recurrent neural networks and attention mechanisms) and graph structure modeling technologies (i.e., graph neural networks) in CDSR, respectively, as shown in Figure~\ref{fig:framework_example}.
Moreover, from Figure~\ref{fig:framework}, we can observe that the three basic technologies are not independent entities and can be applied interchangeably or in parallel based on the characteristics of different technologies. 
Notice that multi-layer perceptron (MLP) is used in almost all existing models for non-linear feature mapping or to simply aggregate cross-domain information~\cite{AMID,MSECDR}, so we do not describe it separately.

\subsubsection{Recurrent Neural Networks in CDSR}
 Recurrent neural networks (RNNs) model sequences, by taking the current time step's representation and the hidden features from the previous time step as input, and generating the output and hidden features for the current time step.
 Due to the issues of vanishing and exploding gradients, its variants, i.e., gated recurrent unit (GRU) and long short-term memory (LSTM)~\cite{SCLSTM}, are more widely adopted.
In CDSR, some works~\cite{DASL,DAT-MDI,CMVCDR} directly employ GRUs as encoders of sequences to get sequential representations.
While some works incorporate a shared unit into each step of RNNs to transfer cross-domain information.
The shared unit could be a shared-account filter unit~\cite{pi-Net,PSJNet} or the common representations of the overlapping users~\cite{CDHRM}.
While RNNs are easy to implement and are capable of modeling the temporal relationships in sequences, they still suffer from issues such as vanishing and exploding gradients during training.

\subsubsection{Attention Mechanisms in CDSR}
Attention mechanisms can selectively focus on the importance of different parts and weight their contributions accordingly, enabling effective aggregation.
Attention mechanisms are utilized in two main ways in CDSR. 
One is to use an attention-based encoder (e.g., Transformer~\cite{Transformer}, SASRec~\cite{SASRec} or multi-head attention blocks, etc.), replacing  RNNs as a sequence encoder~\cite{DREAM,MGCL,CD-SASRec,TPUF,CGRec}. 
The other is to obtain learnable attention weights thus aggregating cross-domain information at multiple levels.
For instance, MiNet~\cite{MiNet} designs item-level attention and interest-level attention to learn which items are more important and which of these items better matches a certain interest of the user.
With representation-level attention, C$^2$DSR~\cite{C2DSR} fuses the sequential representation and the graph representation together.
By adopting domain-level attention, Tri-CDR~\cite{Tri-CDR} learns the weights of the features from different domains and then aggregates them.
And MAN~\cite{MAN} also proposes a group-level attention to bridge the not-aligned information.
In fact, attention strategies need to be designed from multiple levels which poses a challenge to researchers.
However, it needs to use the interaction of all the positions when calculating the weights, so it contains a large number of parameters and may suffer from the issue of data sparsity.
\subsubsection{Graph Neural Networks in CDSR}
Aware of the structural relationships of item-item and item-user in a sequence, graph neural networks are used to model such information.
Some works construct a graph per session~\cite{DAT-MDI} or per domain~\cite{C2DSR,MGCL}, or just construct a global graph~\cite{DA-GCN,LEA-GCN} based on the hybrid domain of all users.
Apart from that, DCDIR~\cite{DCDIR} and MIFN~\cite{MIFN} construct a knowledge graph to encompass more semantic information.
Most works combine graph encoders (i.e., GNNs) and sequence encoders (i.e., RNNs or attention mechanisms) as complementary parts in CDSR.
For instance, DA-GCN~\cite{DA-GCN} utilizes graph convolutional networks to learn latent user representations and user-specific item representations and then carries the weights from the item and user neighbors to the target item in each domain with an attention matrix.
C$^2$DSR~\cite{C2DSR} combines the encoded representations from GNNs with the original sequences and feeds them into the attention encoder for further modeling.
It supplementally learns complex information about nodes and edges to capture more comprehensive preferences of users. 
However, when GNNs are applied to a large-scale data,  the huge computational and storage overhead is a major drawback in most cases.

\begin{table*}[t]
\centering
\caption{A summary of commonly used datasets for CDSR.}
\label{tab:datasets}
\resizebox{1\linewidth}{!}{
\begin{tabular}{c|cc|c|c|c}
\hline
Datasets & \multicolumn{2}{c|}{Domains} & Data types & Scale & Link \\ \hline
\multirow{2}{*}{\begin{tabular}[c]{@{}c@{}}HVIDEO \\ \cite{pi-Net}\end{tabular}} & \multicolumn{2}{c|}{\begin{tabular}[c]{@{}c@{}}V-domain:\\family videos\end{tabular}} & \multirow{2}{*}{user ID, item ID, time} & \multirow{2}{*}{0.4 million +} & \multirow{2}{*}{\tiny\url{https://bitbucket.org/Catherine_Ma/pinet_sigir2019/src/master/HVIDEO}} \\ \cline{2-3}
 & \multicolumn{2}{c|}{\begin{tabular}[c]{@{}c@{}}E-domain:\\educational videos\end{tabular}} &  &  &  \\ \hline
\multirow{3}{*}{\begin{tabular}[c]{@{}c@{}}Douban \\ \cite{GA-DTCDR}\end{tabular}}& \multicolumn{2}{c|}{Movies} & \multirow{3}{*}{\begin{tabular}[c]{@{}c@{}}user ID, item ID, ratings,\\labels, reviews, time, users contexts\end{tabular}} & \multirow{3}{*}{1 millions +} & \multirow{3}{*}{\tiny\url{https://github.com/FengZhu-Joey/GA-DTCDR/tree/main/Data}} \\ \cline{2-3}
 & \multicolumn{2}{c|}{Musics} &  &  &  \\ \cline{2-3}
 & \multicolumn{2}{c|}{Books} &  &  &  \\ \hline
\multirow{5}{*}{\begin{tabular}[c]{@{}c@{}}Amazon \\ \cite{Amazon}\end{tabular}} & \multicolumn{2}{c|}{Books} & \multirow{5}{*}{\begin{tabular}[c]{@{}c@{}}user ID, item ID, ratings,\\time, reviews, item contexts\end{tabular}} & \multirow{5}{*}{100 millions +} & \multirow{5}{*}{\tiny\url{https://jmcauley.ucsd.edu/data/amazon}} \\ \cline{2-3}
 & \multicolumn{2}{c|}{Movies} &  &  &  \\ \cline{2-3}
 & \multicolumn{2}{c|}{Foods} &  &  &  \\ \cline{2-3}
 & \multicolumn{2}{c|}{Kitchens} &  &  &  \\ \cline{2-3}
 & \multicolumn{2}{c|}{…} &  &  &  \\ \hline
\multirow{4}{*}{\begin{tabular}[c]{@{}c@{}}Tenrec \\ \cite{Tenrec}\end{tabular}} & \multicolumn{1}{c|}{\multirow{2}{*}{Videos}} & QK-video & \multirow{2}{*}{\begin{tabular}[c]{@{}c@{}}user ID, item ID, multiple-behavior interactions,\\ video category, watching times, user gender, user age\end{tabular}} & \multirow{4}{*}{100 millions +} & \multirow{4}{*}{\tiny\url{https://static.qblv.qq.com/qblv/h5/algo-frontend/tenrec_dataset.html}} \\ \cline{3-3}
 & \multicolumn{1}{c|}{} & QB-video &  &  &  \\ \cline{2-4}
 & \multicolumn{1}{c|}{\multirow{2}{*}{Articles}} & QK-article & \multirow{2}{*}{\begin{tabular}[c]{@{}c@{}}user ID, item ID, multiple-behavior interactions, \\ read percentage, item contexts, read time\end{tabular}} &  &  \\ \cline{3-3}
 & \multicolumn{1}{c|}{} & QB-article &  &  &  \\ \hline
\multicolumn{1}{c|}{\multirow{3}{*}{\begin{tabular}[c]{@{}c@{}}Mybank-CDR\\ \cite{AMID}\end{tabular}}} & \multicolumn{2}{c|}{Loan} & \multicolumn{1}{c|}{\multirow{3}{*}{user ID, interactive sequences}} &\multirow{3}{*}{100 millions +} & \multirow{3}{*}{\tiny\url{https://github.com/WujiangXu/AMID/tree/main/mybank_dataset}}  \\ \cline{2-3}
 & \multicolumn{2}{c|}{Fund} &  & & \\ \cline{2-3}
 & \multicolumn{2}{c|}{Account} &  &  &\\ \hline
\end{tabular}
}
\end{table*}

\begin{table*}[]
\centering
\caption{Experimental results (\%) on two domains of the Foods and Kitchen of Amazon.
Notice that the results are copied from~\protect\cite{C2DSR,P-CDSR,DREAM} for reference. We bold the best results and underline the second-best results.}
\label{tab:results}
\resizebox{1\linewidth}{!}{
\begin{tabular}{ll|cccccccccccc}
\hline
\multirow{3}[6]{*}{} & \multirow{3}[6]{*}{} & \multicolumn{6}{c|}{Foods}                    & \multicolumn{6}{c}{Kitchens} \\
\cline{3-14}          &       & \multicolumn{1}{c|}{MRR} & \multicolumn{2}{c|}{NDCG} & \multicolumn{3}{c|}{HR} & \multicolumn{1}{c|}{MRR} & \multicolumn{2}{c|}{NDCG} & \multicolumn{3}{c}{HR} \\
\cline{3-14}          &       & \multicolumn{1}{c|}{@10} & \multicolumn{1}{c|}{@5} & \multicolumn{1}{c|}{@10} & \multicolumn{1}{c|}{@1} & \multicolumn{1}{c|}{@5} & \multicolumn{1}{c|}{@10} & \multicolumn{1}{c|}{@10} & \multicolumn{1}{c|}{@5} & \multicolumn{1}{c|}{@10} & \multicolumn{1}{c|}{@1} & \multicolumn{1}{c|}{@5} & @10 \\
    \hline
    \multirow{2}[2]{*}{OCCF} & BPRMF~\cite{BPR} & 4.10  & 3.55  & 4.03  & 2.42  & 4.51  & 5.95  & 2.01  & 1.45  & 1.85  & 0.73  & 2.18  & 3.43  \\
          & ItemKNN~\cite{ItemKNN} & 3.92  & 3.51  & 3.97  & 2.41  & 4.59  & 5.98  & 1.89  & 1.28  & 1.75  & 0.58  & 1.99  & 3.26  \\
    \hline
    \multirow{3}[2]{*}{SOCCF} & GRU4Rec~\cite{GRURec} & 5.79  & 5.48  & 6.13  & 3.63  & 7.12  & 9.11  & 3.06  & 2.55  & 3.10  & 1.61  & 3.50  & 5.22  \\
          & SASRec~\cite{SASRec} & 7.30  & 6.90  & 7.79  & 4.73  & 8.92  & 11.68  & 3.79  & 3.35  & 3.93  & 1.92  & 4.78  & 6.62  \\
          & SR-GNN~\cite{SRGNN} & 7.84  & 7.58  & 8.35  & 5.03  & 9.88  & 12.27  & 4.01  & 3.47  & 4.13  & 2.07  & 4.80  & 6.84  \\
    \hline
    \multirow{2}[2]{*}{CD-OCCF} & NCF-MLP~\cite{NCF-MLP} & 4.49  & 3.94  & 4.51  & 2.68  & 5.10  & 6.86  & 2.18  & 1.57  & 2.03  & 0.91  & 2.23  & 3.65  \\
          & CoNet~\cite{CoNet} & 4.13  & 3.61  & 4.14  & 2.42  & 4.77  & 6.35  & 2.17  & 1.50  & 2.11  & 0.95  & 2.07  & 3.71  \\
    \hline
    \multirow{5}[4]{*}{CD-SOCCF (a.k.a. CDSR)} & $\pi$-Net~\cite{pi-Net} & 7.68  & 7.32  & 8.13  & 5.25  & 9.25  & 11.75 & 3.53  & 2.98  & 3.73  & 1.57  & 4.34  & 6.67 \\
          & MIFN~\cite{MIFN} & 8.55  & 8.28  & 9.01  & 6.02  & 10.43 & 12.71 & 4.09  & 3.57  & 4.29  & 2.21  & 4.86  & 7.08 \\
          & C$^2$DSR~\cite{C2DSR} & 8.91  & 8.65  & 9.71  & 5.84  & 11.24 & 14.54 & 4.65  & 4.16  & 4.94  & 2.51  & 5.74  & 8.18 \\
          & P-CDSR~\cite{P-CDSR} & \textbf{9.87} & \underline{9.57} & \underline{10.72} & \textbf{6.66} & \underline{12.34} & \underline{15.94} & \underline{4.78} & \underline{4.37} & \underline{5.08} & \underline{2.69} & \underline{6.06} & \underline{8.27} \\
\cline{3-3}          & DREAM~\cite{DREAM} & \underline{9.33} & \textbf{10.05} & \textbf{11.25} & \underline{6.08} & \textbf{13.75} & \textbf{17.45} & \textbf{4.82} & \textbf{5.19} & \textbf{6.15} & \textbf{2.74} & \textbf{7.52} & \textbf{10.51}\\
    \hline
    \end{tabular}}
\end{table*}
\subsection{Auxiliary Learning}
In addition to the aforementioned key technologies, the utilization of auxiliary learning technologies to facilitate the integration of cross-domain information garners significant attention.
\subsubsection{Transfer Learning}
Transfer learning (TL)~\cite{transferLearning} is primarily employed to transfer knowledge learned from one task to another task.
For instance, CDNST~\cite{CDNST} transfers the novelty-seeking trait learned from a source domain to a target domain.
 SATLR~\cite{SATLR} considers transferring knowledge by multiplying the independently learned representations from one domain with an orthogonal mapping matrix.
Some works adopt dual transfer learning to improve the ability to transfer knowledge.
DAT-MDI~\cite{DAT-MDI} combines a dual transfer model with slot attention to self-adapt item embedding from different domains.
DASL~\cite{DASL} applies a dual embedding component to unify the learning process of user representations and then proposes a dual attention component to incorporate user behaviors in multiple domains.

\subsubsection{Contrastive Learning}
Contrastive learning (CL)~\cite{Contrastive-survey} is also a widely applied technique that leverages the similarities and differences between samples to extract useful information.
In the case of Tri-CDR~\cite{Tri-CDR}, it designs two contrastive learning tasks, i.e., coarse-grained similarity modeling and fine-grained distinction modeling.
Coarse-grained similarity modeling closes three domains' sequence representations of the same user, and fine-grained distinction modeling assumes the distance between domain A and domain B should be larger than the distance between domain B and the hybrid domain.
MGCL~\cite{MGCL} views the local and global item representations of a user as the positive samples and the representations from different users as the negative samples.
Additionally, some works aggregate the sequences and then combine those processed sequences with CL.
DREAM~\cite{DREAM} proposes supervised contrastive learning to minimize the relevance among inter-sequences with different preferences.

\subsubsection{Other Auxiliary Learning Technologies}
In addition to transfer learning and contrastive learning, there are other auxiliary learning technologies. For instance, RecGURU~\cite{RecGURU} and TPUF~\cite{TPUF} train a discriminator until it is unable to distinguish whether a feature belongs to domain A or domain B, thereby achieving the goal of adversarial feature alignment.
FedDCSR~\cite{FedDCSR} leverages federated learning~\cite{Fed_survey} to preserve data privacy.
RL-ISN~\cite{RL-ISN} utilizes rewards in reinforcement learning to determine whether to revise the whole transferred sequence and selects which interactions should be retained, thus alleviating the noise introduced by transferring cross-domain information.
PLCR~\cite{PLCR} treats domain-specific contexts as the prompt and feeds them with domain-agnostic contexts and label features into self-attention blocks to learn prompt embedding.

\subsubsection{Discussion}
Incorporating auxiliary learning technologies aims to enhance a model's ability to capture cross-domain information.
Contrastive learning can make learned representations more discriminative and robust, but it is sensitive to the designed contrastive strategy.
Transfer learning enables the sharing of information across domains, but the effectiveness of knowledge transfer is affected by the correlation between domains, i.e., there is vulnerability to negative transfer~\cite{negativeTransfer-survey}.
Moreover, adversarial learning can unify representations from different domains but its training process is more complex.

\section{Datasets and Experimental Results}
In this section, we summarize a list of commonly used datasets of CDSR, including their corresponding domains, data types, and scales shown in Table~\ref{tab:datasets}.

In order to more fully represent the advantage in CDSR compared to other issues in Table~\ref{tab:introduction} and the performance of models with different technologies in CDSR, we quote the results of representative models from \cite{C2DSR,P-CDSR,DREAM} in Table~\ref{tab:results}. 
Notice that these three papers are consistent in their treatment of this dataset.
The results show that with the increase in information and advancements in methods, the results gradually improve.

\section{Future Directions}
In this section, we provide several promising directions in CDSR for potential developments from different aspects.
\subsubsection{Multi-Domain Simultaneous Improvement}
As mentioned in Section~\ref{sec:inter-domain} and shown in Figure~\ref{fig:structure}, most existing models are primarily established on two domains. 
In real-world applications, users tend to have interactions in multiple domains.
Exploring how to integrate information from multiple domains (e.g., dozens of domains) and simultaneously improve the performance of each domain is a crucial research direction for the future of CDSR.
\subsubsection{Heterogeneous Information Fusion}
Apart from the side information mentioned in Section~\ref{sec:Intra-domain}, there is a large amount of relevant heterogeneous information in real-world applications.
For example, users are likely to transfer from browsing short videos to purchasing items mentioned in the videos.
Therefore, it is worth investigating effective methods that combine heterogeneous information (e.g., image, video, etc.) and traditional ID-based information, to address the challenges in cross-domain recommendation.
Moreover, it is also worth considering introducing heterogeneous sequential behaviors into cross-domain scenarios.
\subsubsection{Deep Utilization of Non-overlapping Information}
Indeed, the majority of current models rely on overlapping users to bridge different domains, but non-overlapping data also contain rich semantic information that is worth exploring and extracting~\cite{IESRec}.
So researchers can conduct deeper studies on non-aligned information.

\subsubsection{Privacy Preservation}
When it comes to sensitive user information, encryption and protection of data is crucial.
Particularly within the realm of CDSR, there is a greater inclusion of user data.
Therefore, designing effective federated learning methods to ensure privacy while minimizing the loss of valuable information in the cross-domain scenario is a meaningful but less studied area~\cite{PPCDSR}.
\subsubsection{Fairness and Interpretability}
Fairness and interpretability are crucial research topics in recommender systems. 
In CDSR, it is essential to reduce the bias between different domains and to interpret cross-domain sequential recommendation results to users.
\subsubsection{More Advanced Technologies}
In Section~\ref{sec:keyTechnologies}, we analyze the technologies used in existing CDSR models from a micro view. 
However, achieving greater leaps in performance demands more advanced technologies.
For instance, exploring the application of large language models (LLMs)~\cite{LLM4Rec} in the CDSR scenario is also a promising direction. 

\section{Conclusions}
Cross-domain sequential recommendation (CDSR) extends traditional recommender systems by incorporating sequential and cross-domain information, aiming to address the data sparsity issue. In this survey, we approach CDSR with a four-dimensional tensor and offer a comprehensive overview from macro and micro views. 
From a macro view, we summarize the existing models by abstracting multi-level fusion structures and discuss the bridges for cross-domain fusion. 
From a micro view, we analyze and summarize the employed basic and auxiliary learning technologies. 
Finally, we include some public datasets and experimental results for CDSR, and provide some promising future research directions.

\section{ACKNOWLEDGMENTS}
We thank the support of National Natural Science Foundation of China (Nos. 62172283 and 62272315), Guangdong Basic and Applied Basic Research Foundation (No. 2024A1515010122) and National Key Research and Development Program of China (No. 2023YFF0725100).
\bibliographystyle{named}
\bibliography{ijcai24}

\end{document}